# Satellite-relayed intercontinental quantum network

Sheng-Kai Liao[1,2], Wen-Qi Cai[1,2], Johannes Handsteiner[3,4], Bo Liu[4,5], Juan Yin[1,2], Liang Zhang[2,6], Dominik Rauch[3,4], Matthias Fink[4], Ji-Gang Ren[1,2], Wei-Yue Liu[1,2], Yang Li[1,2], Qi Shen[1,2], Yuan Cao[1,2], Feng-Zhi Li[1,2], Jian-Feng Wang[7], Yong-Mei Huang[8], Lei Deng[9], Tao Xi[10], Lu Ma[11], Tai Hu[12], Li Li[1,2], Nai-Le Liu[1,2], Franz Koidl[13], Peiyuan Wang[13], Yu-Ao Chen[1,2], Xiang-Bin Wang[2], Michael Steindorfer[13], Georg Kirchner[13], Chao-Yang Lu[1,2], Rong Shu[2,6], Rupert Ursin[3,4], Thomas Scheidl[3,4], Cheng-Zhi Peng[1,2], Jian-Yu Wang[2,6], Anton Zeilinger[3,4], Jian-Wei Pan[1,2]

[1] Hefei National Laboratory for Physical Sciences at the Microscale and Department of Modern Physics, University of Science and Technology of China, Hefei 230026, China.
[2] Chinese Academy of Sciences (CAS) Center for Excellence and Synergetic Innovation Center in Quantum Information and Quantum Physics, University of Science and Technology of China, Shanghai 201315, China
[3] Vienna Center for Quantum Science and Technology, Faculty of Physics, University of Vienna, Vienna 1090, Austria.
[4] Institute for Quantum Optics and Quantum Information, Austrian Academy of Sciences, Vienna 1090, Austria.
[5] School of Computer, National University of Defense Technology, Changsha 410073, China.
[6] Key Laboratory of Space Active Opto-Electronic Technology, Shanghai Institute of Technical Physics, Chinese Academy of Sciences, Shanghai 200083, China.
[7] National Astronomical Observatories, Chinese Academy of Sciences, Beijing 100012, China
[8] Key Laboratory of Optical Engineering, Institute of Optics and Electronics, Chinese Academy of Sciences, Chengdu 610209
[9] Shanghai Engineering Center for Microsatellites, Shanghai 201203, China
[10] State Key Laboratory of Astronautic Dynamics, Xi'an Satellite Control Center, Xi'an 710061, China
[11] Xinjiang Astronomical Observatory, Chinese Academy of Sciences, Urumqi 830011, China
[12] National Space Science Center, Chinese Academy of Sciences, Beijing 100080, China
[13] Space Research Institute, Austrian Academy of Sciences, Graz 8042, Austria.

**Abstract**:

**We perform decoy-state quantum key distribution between a low-Earth-orbit satellite and multiple ground stations located in Xinglong, Nanshan, and Graz, which establish satellite-to-ground secure keys with ~kHz rate per passage of the satellite *Micius* over a ground station. The satellite thus establishes a secure key between itself and, say, Xinglong, and another key between itself and, say, Graz. Then, upon request from the ground command, *Micius* acts as a trusted relay. It performs bitwise exclusive OR operations between the two keys and relays the result to one of the ground stations. That way, a secret key is created between China and Europe at locations separated by 7600 km on Earth. These keys are then used for intercontinental quantum-secured communication. This was on the one hand the transmission of images in a one-time pad configuration from China to Austria as well as from Austria to China. Also, a videoconference was performed between the Austrian Academy of Sciences and the Chinese Academy of Sciences, which also included a 280 km optical ground connection between Xinglong and Beijing. Our work points towards an efficient solution for an ultra-long-distance global quantum network, laying the groundwork for a future quantum internet.**

With the growth of internet use and electronic commerce, a secure global network for data protection is necessary. A drawback of traditional public key cryptography is that it is not possible to guarantee it is information theoretically secure. It has been witnessed in history that every advance of encryption has been defeated by advances in hacking. In particular, with the advent of Shor's factoring algorithm [1], most of the currently used cryptographic infrastructure will be defeated by quantum computers.

On the contrary, quantum key distribution (QKD) [2] offers unconditional security ensured by the law of physics. QKD uses the fundamental unit of light, single photons, encoded in quantum superposition states which are sent to a distant location. By proper encoding and decoding, two distant parties share strings of random bits called secret keys. However, due to photon loss in the channel, the secure QKD distance by direct transmission of the single photons in optical fibers or terrestrial free space was hitherto limited to a few hundred kilometers [3–7]. Unlike classical bits, the quantum signal in the QKD cannot be noiselessly amplified owing to the quantum no-cloning theorem [8], already contained at the core of Wiesner's proposal of uncopiable quantum money [9], where the security of the QKD is rooted.

The main challenge for a practical QKD is to extend the communication range to long distances, ultimately on a global scale. A promising solution to this problem is exploiting satellite and space-based links [10,11]. That way, one can conveniently connect two remote points on Earth with greatly reduced channel loss because most of the photons' propagation path is in empty space with negligible loss and decoherence. Very recently, QKD from a low-Earth-orbit satellite, *Micius*, to the Xinglong ground station close to Beijing has been demonstrated with a satellite-to-ground-station distance of up to 1200 km [12].

In this Letter, we use the *Micius* satellite as a trusted relay to distribute secure keys between multiple distant locations in China and Europe. The *Micius* satellite, launched in August 2016, orbits at an altitude of about 500 km. As illustrated in Fig. 1, the three cooperating ground stations are located in Xinglong (near Beijing, 40°23'45.12"N, 117°34'38.85"E, altitude of 890 m), Nanshan (near Urumqi, 43°28'31.66"N, 87°10'36.07"E, altitude of 2028 m), and Graz (47°4'1.72"N, 15°29'35.92"E, altitude of 490m). The distances from Xinglong to Nanshan and Graz are 2500 km and 7600 km, respectively.

In this work, QKD is performed in a downlink scenario—from the satellite to the ground. One of the payloads in the satellite is a space-qualified QKD transmitter [12], which uses weak coherent laser pulses to implement a decoy-state Bennett-Brassard 1984 (BB84) protocol that is immune to the photon-number-splitting attack [13,14]. Eight tunable fiber lasers, emitting light pulses with a wavelength of ~850 nm at a repetition rate of 100 MHz, are used to generate the signal, decoy and vacuum states. After being collected into single-mode fibers and collimated, the laser pulses enter a BB84-encoding module. It consists of a half-wave plate (HWP), two polarizing beam splitters (PBSs), and one non-polarizing beam splitter (BS). The photons emitted and sent to the ground station are randomly prepared in one of the four polarization states: horizontal, vertical, linear 45°, and linear -45°. In the three ground stations, corresponding BB84-decoding setups are used, consisting of a BS, a HWP, two PBS and four single-photon detectors (see Supplemental Material).

For secure QKD, the average intensity per pulse sent over the channel has to be at the single-photon level. As the photons travel from the fast-moving satellite (~7.6 km/s) through the atmosphere to the ground station over typically ~1000 km, several effects contribute to the channel loss such as beam diffraction, pointing errors, atmospheric turbulence and absorption. As is typical for photonic communication, decoherence can be ignored. To obtain a high signal-to-noise ratio in the QKD protocol, one cannot increase the signal power but only reduce the channel attenuation and background noise. In order to optimize the link efficiency, we combine a narrow transmitting beam divergence (~10 μrad) with high-bandwidth acquisition, pointing, and tracking technique that ensures a typical tracking accuracy of ~1 μrad (ref. [12] and Extended Data Table I). To reduce the background noise, the BB84-decoding setups in the optical ground stations are designed with a small field-of-view and employ low dark-count rate single-photon detectors.

The satellite flies along a sun-synchronized orbit which circles Earth every 94 minutes. Each night starting at around 0:50 AM local time, *Micius* passes over the three ground stations allowing for a downlink for a duration of ~300 s. Under reasonably good weather conditions, we can routinely obtain a sifted key rate of a ~3 kb/s at ~1000 km physical separation distance and ~9 kb/s at ~600 km distance (at the maximal elevation angle, note that this distance is greater than the satellite height due to the fact that the satellite does not fly directly over the ground stations), respectively. The observed quantum bit error rates are in the range of 1.0%-2.4%, which is caused by background noise and polarization errors. The satellite is equipped with an experimental control box (see Supplemental Material) that is able to exchange classical data with dedicated ground stations through radio frequency channels, with uplink and downlink bandwidth of 1 Mbps and 4 Mbps, respectively. This allows us to implement the full QKD protocol including sifting, error correction and privacy amplification (see Supplemental Material), to obtain the final keys between the satellite and the three ground stations. Typical satellite-to-ground QKD performances between May and July 2017 are summarized in Figure 1, with the final key length ranging from 400 kb to 833 kb.

Next, we rely on the satellite as a trusted relay to establish secure keys among the ground stations on Earth. Figure 2 shows the example of exchanging keys between the Xinglong and Graz stations. We denote the random keys shared between *Micius* and Xinglong as $MX$, and between *Micius* and Graz as $MG$. *Micius* can simply perform a bitwise exclusive OR operation ($\oplus$) between $MX$ and $MG$ of the same string length, which then yields a new string: $MX \oplus MG$. The new string can then be sent through a classical communications channel to Xinglong or Graz, who can decode the other's original key by another exclusive OR (i.e. $MG = (MX \oplus MG) \oplus MX$). This process can be easily understood as *Micius* uses *MX* to encrypt *MG* and Xinglong decrypts the cipher text to recover *MG*, shared with Graz. Such a key is known only to the two communicating parties and the satellite, but not any fourth party. In this work, we establish a 100 kB secure key between Xinglong and Graz. Similarly, secure keys between Nanshan and Xinglong, and between Nanshan and Graz can also be established.

For a real-world application of the space-to-ground integrated quantum network, we transmitted a picture of Micius (with a size of 5.34 kB) from Beijing to Vienna, and a picture of Schrödinger (with a size of 4.9 kB) from Vienna to Beijing, using approximately 10kB from the 100kB secure quantum key for one-time-pad encoding (Figure 3).

On 29th of September 2017, an intercontinental video conference was held between the Chinese Academy of Sciences and the Austria Academy of Sciences. The satellite-based QKD network is combined with fiber-based metropolitan quantum networks, in which fibers are used to efficiently and conveniently connect many users inside a city with a distance scale of within 100 km. The Xinglong ground station is connected to the conference venue, Zhongguancun Software Park in Beijing via a 280-km optical fiber link involving six trusted relays [15,16]. We employed the Advanced Encryption Standard (AES)-128 protocol that refreshed the 128-bit seed keys every second. The video conference lasted for 75 minutes with a total data transmission of ~2 GB, which consumed ~70 kB of the quantum key generated between Austria and China.

In summary, using *Micius* satellite as a trusted relay, we have demonstrated intercontinental quantum communication among multiple locations on Earth with a maximal separation of 7600 km. Our work already constitutes a simple prototype for a global quantum communications network. To increase the time and area coverage for a more efficient QKD network, we plan to launch higher-orbit satellites and implement day-time operation using telecommunication wavelength photons and tighter spatial and spectral filtering [17]. One limitation of the current implementation of the QKD protocol is that we have to trust the satellite itself, which can be overcome in the future using entanglement-based systems [18–20]. Other future developments will include multi-party connections from satellites to various ground stations in parallel, and the connection to large ground networks [16], at first in China and Europe and then on a global scale.

**Figure caption**

Figure 1: Illustration of the three cooperating ground stations (Graz, Nanshan and Xinglong). Listed are all paths which were used for key generation and the corresponding final key length.

Figure 2: Schematic of key exchange procedure between Graz and Xinglong with the satellite as a trusted relay. After Micius distributed a key with Graz (MG) and Xinglong (MX) it performs a bitwise exclusive OR between those keys ( $MX \oplus MG$ ) and sends this combined key via a classical channel towards Xinglong station. Combining the XORed key at Xinglong with MX leads to the same key (MG) on both sides.

Figure 3: One-time-pad file transfer. A picture of Micius/Schrödinger was transferred between Beijing and Vienna one-time-pad encrypted with a secure key with a length of 5.34kB/4.9kB, ensuring unconditional security. Binary view of pictures and keys are depicted where each pixel represents one byte of data/key encoded in a 256 color scale. Each side en-/decrypted the picture with a simple bitwise XOR operation.

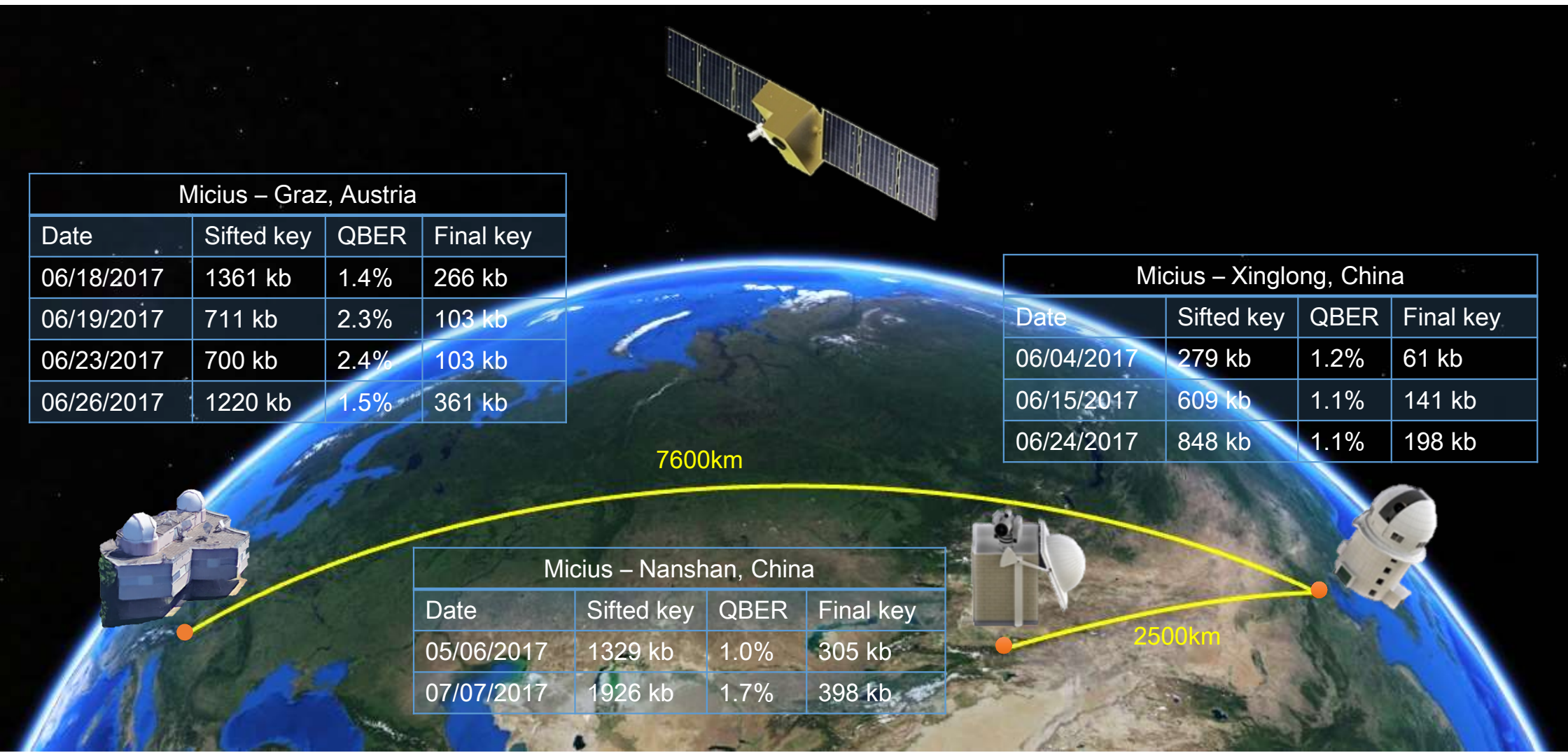

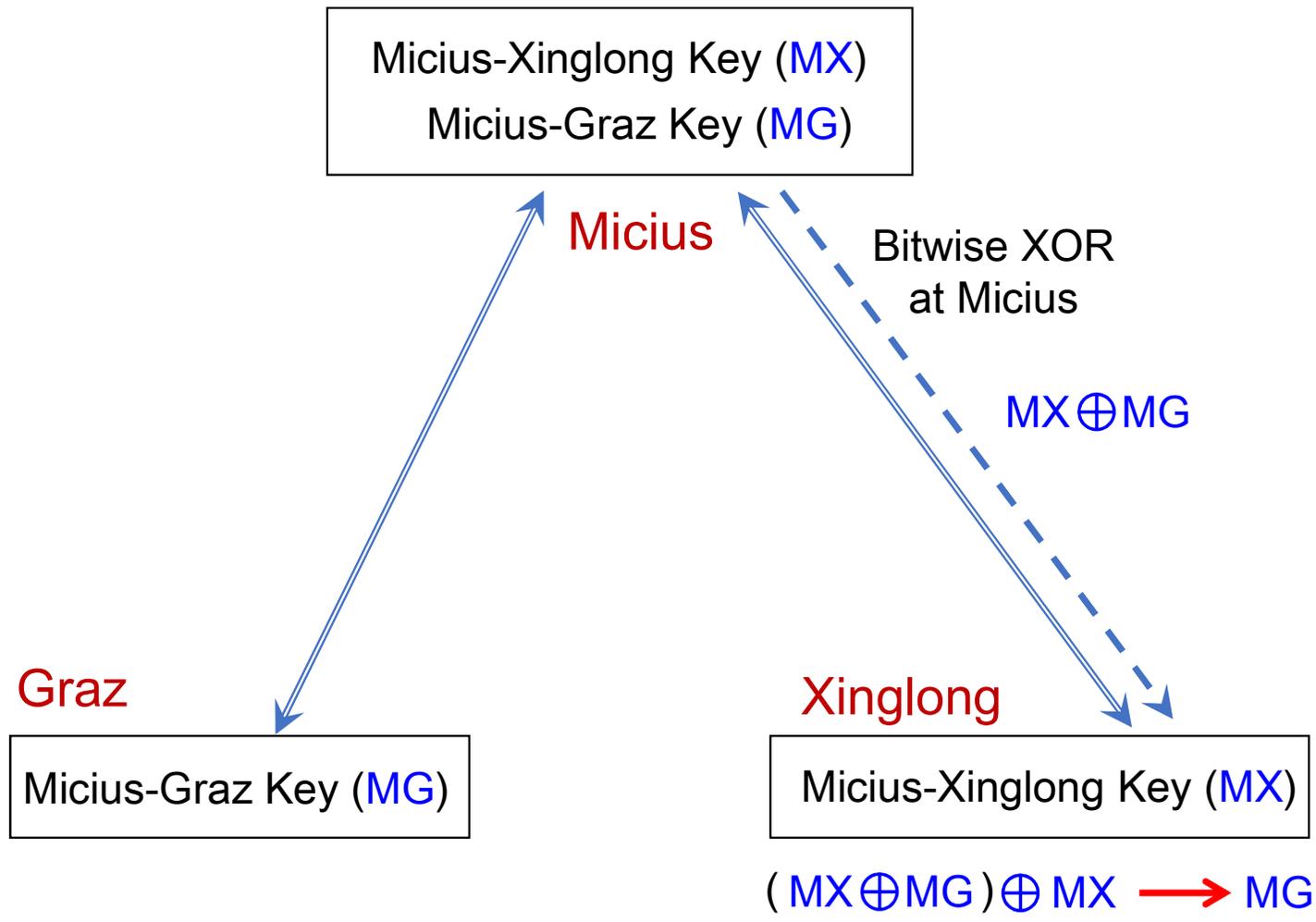

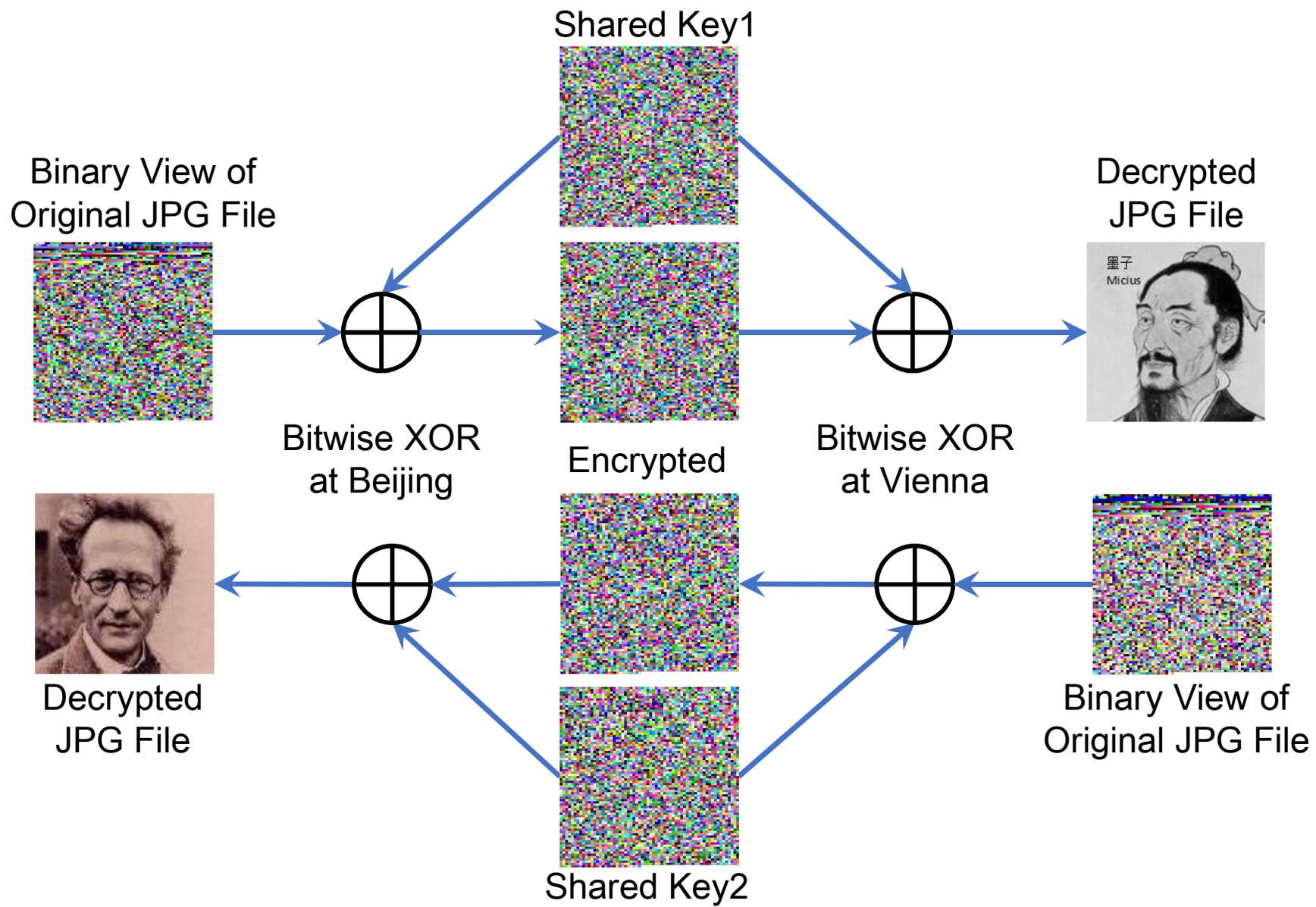